\documentclass[letterpaper]{article} % DO NOT CHANGE THIS
\usepackage{aaai24}  % DO NOT CHANGE THIS
\usepackage{times}  % DO NOT CHANGE THIS
\usepackage{helvet}  % DO NOT CHANGE THIS
\usepackage{courier}  % DO NOT CHANGE THIS
\usepackage{amsmath}
\usepackage{subcaption}
\usepackage[algo2e]{algorithm2e} 
\usepackage[hyphens]{url}  % DO NOT CHANGE THIS
\usepackage{graphicx} % DO NOT CHANGE THIS
\usepackage{natbib}  % DO NOT CHANGE THIS AND DO NOT ADD ANY OPTIONS TO IT
\usepackage{caption} % DO NOT CHANGE THIS AND DO NOT ADD ANY OPTIONS TO IT
\usepackage{algorithm}
\usepackage{newfloat}
\usepackage{listings}
\DeclareCaptionStyle{ruled}{labelfont=normalfont,labelsep=colon,strut=off} % DO NOT CHANGE THIS
\floatstyle{ruled}
\newfloat{listing}{tb}{lst}{}
\floatname{listing}{Listing}
\title{On the Potential of Network-Based Features for Fraud Detection}
\author {
    Catayoun Azarm\textsuperscript{\rm 1},
    Erman Acar\textsuperscript{\rm 1},
    Mickey van Zeelt\textsuperscript{\rm 2}
}
\affiliations {
    \textsuperscript{\rm 1}University of Amsterdam\\
    \textsuperscript{\rm 2}ING Netherlands\\
    catayounazarm@yahoo.de, e.acar@uva.nl, mickey.van.zeelt@ing.com
}

\usepackage{bibentry}
\begin{document}

\maketitle

\begin{abstract}
Online transaction fraud presents substantial challenges to businesses and consumers, risking significant financial losses. Conventional rule-based systems struggle to keep pace with evolving fraud tactics, leading to high false positive rates and missed detections. Machine learning techniques offer a promising solution by leveraging historical data to identify fraudulent patterns. This article explores using the personalised PageRank (PPR) algorithm to capture the social dynamics of fraud by analysing relationships between financial accounts. The primary objective is to compare the performance of traditional features with the addition of PPR in fraud detection models. Results indicate that integrating PPR enhances the model's predictive power, surpassing the baseline model. Additionally, the PPR feature provides unique and valuable information, evidenced by its high feature importance score. Feature stability analysis confirms consistent feature distributions across training and test datasets.
\end{abstract}

\section{Introduction}

\label{sec:introduction}

Online transaction fraud detection (FD) is a significant challenge due to its widespread occurrence and potential for financial losses \cite{carcillo2018beyond}. Traditional rule-based systems struggle with evolving fraud tactics, leading to high false positives and missed detections \cite{zhang2018model}.

Machine learning offers promise in addressing this issue, using historical data to identify patterns and anomalies signaling fraudulent behaviour \cite{chauhan2020fraud}. Accurate and reliable FD models are crucial to effectively mitigate risks linked to online transaction fraud.

However, fraudulent activities involve a social dimension, extending beyond individual incidents. Recent research emphasises considering the relational aspects of fraud. Innovative techniques leveraging intricate relationships between financial accounts are crucial in combating sophisticated fraudulent activities, with graph networks providing a holistic perspective and aiding in identifying intricate patterns \cite{neo4j}.

An example involves applying the personalised PageRank algorithm to a credit card transaction network, as introduced in \cite{van2015apate}. The algorithm uses collective inference to calculate exposure scores for each node, reflecting the influence's propagation from confirmed fraudulent transaction nodes. These scores indicate a network object's exposure to fraud.

Integrating network-based exposure features is crucial, offering valuable insights into relationships and interconnections among transactional entities. These features quantify node significance relative to a specific user or group. Incorporating them into FD models can also help with identifying suspicious nodes and uncovering hidden connections. 

This paper aims to compare baseline features against the personalised PageRank exposure score and elucidate their respective contributions. Key contributions of this research lie in analysing the efficacy of personalised PageRank exposure scores, particularly in comparison to conventional baseline features, in enriching FD models. This includes evaluating their impact on model performance, interpretability, and feature stability. Furthermore, this research aims to validate the effectiveness of integrating the personalised PageRank exposure feature with baseline features in enhancing overall FD capabilities. To achieve this, a transaction graph is created, and personalised PageRank is computed to capture intricate relationships between current and past transactions. This exposure feature is integrated with baseline features to evaluate the combined feature set's effectiveness.\footnote{The source code is available at \url{https://github.com/Catayoun/ppr_paper}.}  

\section{Related Work}
\label{sec:related_work}

\subsection{Financial Fraud Detection}

Extensive research has focused on enhancing Financial Fraud Detection (FFD) accuracy by employing techniques such as machine learning, data mining, and rule-based analysis. This section delves into these commonly used methods and contextualises the approach within existing frameworks.

\subsubsection{Machine Learning}

Machine learning has been a focal point in FFD studies, with significant emphasis on feature selection and the construction of supervised classifiers. Approaches include Bayesian networks (BN) \cite{kirkos2007data}, decision trees \cite{shen2007application}, Genetic Algorithms \cite{duman2013solving}, and Support Vector Machines (SVM) \cite{chen2006new}.

\subsubsection{Data Mining}

Data mining plays a crucial role in FFD, extracting valuable insights and revealing hidden patterns in extensive datasets \cite{ngai2011application}. Key data mining techniques include Logistic Models, BNs \cite{bermudez2008bayesian}, and NNs \cite{kirkos2007data}.

For instance, \cite{nami2018cost} introduced a two-stage method using Random Forest and K-Nearest Neighbor (KNN) algorithms for payment card FD. \cite{ganji2012credit} proposed a reverse KNN algorithm for credit card fraud detection (CCFD). In \cite{alimolaei2015intelligent}, a system was designed to detect abnormal user behaviour in Internet banking. \cite{save2017novel} employed supervised learning methods, integrating a decision tree algorithm with other CCFD techniques. \cite{zanin2018credit} suggested a hybrid approach combining data mining and complex network classification algorithms for CCFD. \cite{abdulla2015hybrid} developed a FD system using a Genetic Algorithm for feature selection and SVMs for classification.

\subsubsection{Rule-Based Approach}

The rule-based FD approach assesses absolute and differential usage against predefined rules, offering flexibility through differential analysis to detect changes in usage patterns based on user behaviour history. Rule-based analysis prefers user profiles with explicit information, allowing fraud criteria to be defined as rules.

A rule-discovery method proposed by \cite{rosset1999discovery} combines customer and behavioural data, using a greedy algorithm with adjusted thresholds. \cite{rajput2014ontology} introduced an ontology-based system for detecting fraudulent transactions, employing an ontology graph-based approach. 

Despite its benefits, rule-based analysis presents challenges, requiring precise and time-consuming programming for rule configuration. Ongoing adaptation is necessary for emerging fraud types, and scalability becomes an issue as system performance degrades with increased data volume \cite{kou2004survey}.

\subsection{Network-Based Fraud Detection}

The banking industry faces challenges in detecting interconnected and collusive fraud schemes. Traditional methods are often insufficient, requiring advanced approaches. Network-based FD analyses transactions within interconnected networks, revealing hidden connections and dependencies \cite{al2020critical}. This method comprehensively explains fraudulent behaviour through network analysis and graph theory.

\subsubsection{Graph Analytics}

Graph analytics has gained popularity for anomaly detection in financial transaction networks. \cite{noble2003graph} proposed using variants of the Minimum Description Length (MDL) principle to identify anomalous sub-graphs, while \cite{akoglu2010oddball} developed an approach to detect anomalous nodes in large graphs. APATE \cite{van2015apate} is a pioneering study leveraging network information for FD, integrating graph-based features connecting credit cards and merchants with traditional features. APATE employs an extensive feature engineering process, including extracting personalised PageRank scores from the transaction network, which are then used as input for a downstream classifier.

\subsubsection{Neural Networks}

In general, neural networks have found application in multiple products and publications to assess transactions based on statistical attributes \cite{ghosh1994credit}. Neural networks, emulating the human brain's functionality with interconnected nodes, show promise in detecting credit card fraud \cite{chaudhary2012review}. \citet{ghosh1994credit} introduced an approach using neural networks for FD. 
%By learning from genuine cardholders' past actions, neural networks establish standard behaviour patterns, identifying anomalies and preventing fraud. 
These approaches exhibit higher detection rates and fewer false positives compared to other methods.

\subsubsection{PageRank Algorithm}

PageRank, a widely used algorithm in link analysis \cite{brin1998anatomy}, determines node importance in a graph by assessing connections \cite{molloy2017graph}. It calculates probabilities, signifying a node's likelihood of being reached by a random walker, with higher probabilities denoting greater importance.

The algorithm initiates with a random walker from a chosen node, navigating through edges or teleporting, and iteratively computes PageRank values representing probabilities. Significantly, the personalised PageRank (PPR) algorithm begins from specific seed nodes, leading to biased graph exploration. In generating an exposure feature that highlights node relevance, PPR becomes instrumental in FD by discerning anomalous transactions \cite{haveliwala2002topic}. In this context, this paper leverages PPR to detect anomalies and unusual graph behaviour by prioritising specific nodes for scrutiny.

The correlation between PageRank values and transaction frauds lies in their reflection of network centrality. Higher PageRank values can indicate potential fraud by emphasising nodes crucial to biased graph exploration, revealing unexpected patterns. Conversely, lower PageRank values may suggest less relevance, potentially signaling abnormal transactions that deviate from established patterns. This strategic application of graph-based features, exemplified by PageRank, enhances machine learning algorithms for FD, ultimately reducing false positives \cite{molloy2017graph}. In transaction graphs, where nodes represent bank accounts and edges signify unique transactions, the incorporation of PageRank offers an effective means of reassessing fraud classifications based on the network structure.

\begin{figure}[H]
  %\centering
  \includegraphics[width=235pt]{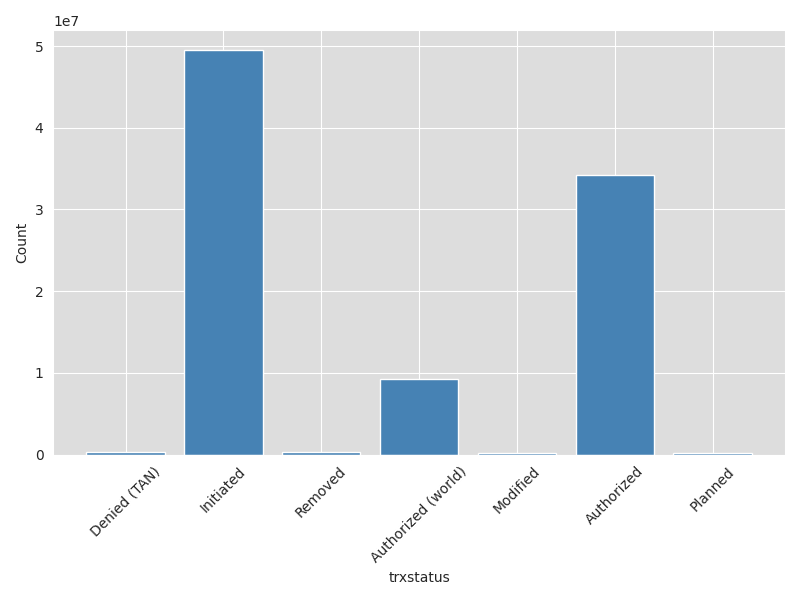}
  \caption{Transaction Status Distribution.}
  \label{fig:trxstatus}
\end{figure}

\section{Methodology}

\label{sec:methodology}

\subsection{Experimental Setup}

To evaluate the proposed approach, a unique dataset consisting of approximately 89 million transactions from ING Netherlands online transaction data was used. The dataset captures a duration of two months, precisely spanning from September 2020 to October 2020. If a transaction is identified as fraudulent, it is marked as "1", otherwise, it is marked as "0". The dataset comprises 15 columns, see Table \ref{table:columns-table}.

During data cleansing and pre-processing, a notable observation emerges regarding the transaction status within the dataset. Specifically, a significant proportion of transactions are classified as "Initiated", indicating their considerable significance within the dataset (Figure \ref{fig:trxstatus}). This early-stage status strategically offers an opportunity for focused FD, distinct from completed transactions. This methodological choice refines the focus on the initial phases of transactional activity, identifying potential threats during initiation before further processing. Consequently, this paper exclusively examines transactions with the "Initiated" status, which constitute 47\% of the entire dataset. This intentional selection facilitates a concentrated analysis of the most relevant subset, while the remaining dataset of 49 million transactions provides ample data for a comprehensive exploration of FD strategies. Importantly, the dataset reveals a significant class imbalance, with only 3198 instances of fraud, representing approximately 0.00006\% of the transactions. 

After feature engineering, the dataset encompasses six distinct features (current\_amount, current\_amount\_first\_digit, channel\_index, trx\_count\_creditor, day \_of\_week, time\_of\_day) carefully chosen based on ING's baseline features, which have demonstrated effectiveness in FD tasks. The baseline features capture various aspects of the transaction data, including static and time window features. Additionally, the network-based feature \emph{ppr} is incorporated, which offers insights into the network structure and its influence on FD. This feature considers the relationships between entities in the transaction network and their respective importance. To improve FD accuracy, logistic regression is chosen as the primary approach. Widely recognised and effective for binary classification tasks, logistic regression provides interpretability, efficiency, and scalability to handle large datasets with multiple features \cite{nick2007logistic}.

The final training set is formed with 70\% of the total transaction data (about 34 million transactions), while the remaining 30\% forms the test set (approximately 15 million transactions). Further sections will provide detailed insights into these features, highlighting their significance and influence on the model's ability to detect fraudulent behaviour.

\begin{table}[]
\footnotesize
\centering
\begin{tabular}{ |p{4.8cm}|p{0.5cm}|p{2.3cm}|  }
 \hline
 \multicolumn{3}{|c|}{\textbf{Overview of columns}} \\
 \hline
 \textbf{Column name} &  \textbf{Type} & \textbf{Example}\\
 \hline
 \texttt{id}           &  str    & "342004885"\\
 \texttt{eventtimecet}                 &  time    & 2020-08-29T16:32:59 \\ 
 \texttt{trxamount}             &  float    & 721.21\\
 \texttt{currency}                     &  str & "EUR" \\
 \texttt{transactionexecutiondate}            &  date    & 
2020-08-29 \\
 \texttt{transactiontype}            &  str & "Domestic"\\
 \texttt{trxstatus
}         &  str    & "Initiated" \\
 \texttt{channel}               &  str    & "DIRECT\_WEB"\\
 \texttt{label}              &  int   & 0\\
 \texttt{debtoraccountnumberhash}         &  int   & 10315208 \\
\texttt{creditoraccountnumberhash}              &  int   & 19209040 \\
 \texttt{partyidhash}              &  int   & 5487856 \\ 
 \texttt{sourceiphash}             &  int   & 5137061\\
 \texttt{sessionidhash}           &  int   & -8869831554710\\
 \texttt{creditoraccountpartyidhash}             &  int   & 5233759\\ 
 \hline
 \end{tabular}
\caption{Overview of the columns in the dataset, their datatype and an example value.} 
\label{table:columns-table} 
\end{table}

\subsection{Baseline Model Features}

The following section provides an overview of features in the baseline logistic regression model. Classified as static and time window features, these features are derived from the ING dataset, offering insights into transaction behaviour.

\subsubsection{Static Features}

The static features encompass vital details about the transaction and the entities involved, offering a momentary representation of transaction attributes and party characteristics. They provide a snapshot at a specific moment in time without considering temporal dynamics or historical patterns. Table \ref{table:static} summarises the static features and gives an example. Incorporating these features enhances the model's ability to understand and analyse the immediate characteristics of a transaction. The following features are extracted: \\

\textbf{Current Amount.} The numerical value of the transaction at a specific point in time.

\textbf{Current Amount First Digit.} The first digit of the transaction amount. It provides information about the leading digit of the transaction amount, such as 1, 2, 3, etc. For instance, if a transaction amount is 250.0, the first digit is 2.

\textbf{Channel Index.} The indexed channel through which the transaction occurred. It indicates the platform used to conduct the transaction, such as web or mobile retail.

\begin{table}[H]  
\footnotesize
\centering
\begin{tabular}{| l |p{0.6cm}| p{1.3cm} |}  
\hline
 \multicolumn{3}{|c|}{\textbf{Overview of features}} \\
 \hline
\textbf{Feature name} & \textbf{Type} & \textbf{Example} \\ \hline 
\texttt{current\_amount} &  float & 250.0  \\   
\texttt{current\_amount\_first\_digit} & int &   1\\
\texttt{channel\_index} & float & 0.2  \\ \hline
\end{tabular}  
\caption{Overview of Static Features} 
\label{table:static}
\end{table}

\subsubsection{Time Window Features}

In addition to static features, time window features capture temporal dynamics and historical patterns by considering the specific moment in time and surrounding transaction history. Using a sliding time window, the model covers a period of 1.5 months during data development, reserving the remaining two weeks exclusively for building historical information. This approach allows a thorough analysis of previous transactions within a specific timeframe, enabling the model to capture patterns and trends over time and providing valuable insights into transactional temporal dynamics. The following features are extracted (see Table \ref{table:twf}) \\

\textbf{Transaction count creditor.} The number of external transactions to a specific creditor within a particular time window, considering the historical transaction history. It counts the total number of transactions that involve the creditor as the receiving party. 

\textbf{Day of week.}  Comparison of the day of the transaction with the debtor's historical patterns to determine if it aligns with their typical transaction behaviour. Examining the consistency of transaction days, one can detect any deviations. 

\textbf{Time of day.} Comparison of the average amount of transactions made by a specific customer. It calculates the average transaction amount for the customer over the past seven days and compares it to the average amount over a more extended historical period.

\begin{table}[H]     
\footnotesize
\centering
\begin{tabular}{| l |p{0.6cm}| p{2cm} |}  
\hline
 \multicolumn{3}{|c|}{\textbf{Overview of features}} \\
 \hline
\textbf{Feature name} & \textbf{Type} & \textbf{Example} \\ \hline  
\texttt{trx\_count\_creditor} & int &  25 \\
%\texttt{trx\_count\_debtor} & int &  8 \\
\texttt{day\_of\_week} & float & 0.7  \\   
\texttt{time\_of\_day} & float & 0.7  \\ \hline
\end{tabular}  
\caption{Overview of the Time Window Features.} \label{table:twf} 
\end{table}

\subsection{Network Feature Extraction}

\subsubsection{Network Generation}

To compute PPR values, a transaction graph $G$ is constructed. This pre-built graph, denoted as $G=(V,E)$ represents relationships between bank accounts and carries weights based on specific properties. $G$ is a unipartite, directed, and weighted graph consisting of vertices or nodes $v \in V$, connected by a set of edges $e \in E$. To build $G$, individual transactions within a specified timeframe are considered. This allows for a fine-grained analysis of transaction patterns and relationships between accounts. Incorporating transaction-level details directly enriches the resulting graph, capturing the intricate network structure.

The generation of this graph $G$ is achieved through the implementation of the function presented in Algorithm \ref{alg:graph}. The function starts by extracting the relevant columns from the $train\_data$ and converting them into a list of tuples representing the graph's edges. Using the TupleList constructor, a graph is then created with the following attributes as edges:

\begin{itemize}
\item $"trxamount"$: The amount transferred between two accounts in a single transaction.
\item $"id"$: The unique identifier of each transaction.
\end{itemize}

While the graph and its properties have shown promise in similar network analysis projects at ING, it is essential to acknowledge the uncertainty surrounding their optimisation for the specific problem at hand. The applicability and effectiveness of this pre-built network structure in the current research context may not be guaranteed. However, considering the similarities between the use case and the previous scenarios where this network was used, it is deemed an appropriate starting point for the research, allowing the avoidance of extensive optimisation steps. It is crucial to approach the analysis and interpretation of the results cautiously, understanding the potential limitations and trade-offs w.r.t. the network's configuration. The selection of this pre-built network is justified by its alignment with similar use cases and data. This decision enables the insights gained from prior experiences to be leveraged and adapted to the current one.

\begin{algorithm}
  \caption{Graph Generator}\label{alg:graph}
  \DontPrintSemicolon
  \SetAlgoNlRelativeSize{-1}
  
  \SetKwFunction{generateTrxGraph}{generate\_trx\_graph}
  \SetKwProg{Fn}{Function}{:}{}
  
  \Fn{\generateTrxGraph{$train\_data, \\ is\_directed=True$}}{
    \textit{edges} $\gets$ list with relevant fields: [$"id", "debtoraccountnumberhash", \\ "creditoraccountnumberhash", "trxamount"$]\;
    
    \textit{graph} $\gets$ Generate the graph using \textit{ig.Graph} library\;
    \Indp
    Pass the \textit{edges} list as the first argument\;
    Set \textit{directed=True}\;
    Specify the vertex name attribute as "name"\;
    Define the edge attributes as ["trxamount", "id"]\;
    \Indm
    
    \textbf{return} \ \textit{graph}\;
  }
\end{algorithm}

\subsubsection{Feature Extraction}

In the following, the process of feature extraction for computing PPR values is presented. The provided code, as shown in Algorithm \ref{alg:ppr}, undertakes the computation of PPR values based on the given training data. This process involves calculating average label values for creditors and debtors, representing the likelihood or degree of fraudulent activity, with values ranging from 0 to 1. These computed values are subsequently stored in a personalisation dictionary, which will be employed as an input parameter when computing the PPR values. It dictates the probability associated with selecting each node as the source vertex.

To complete the calculation of PPR values, the algorithm uses the $nx.pagerank$ function, which leverages the previously constructed graph. The damping factor, denoted as $\alpha=0.85$, is employed to regulate the behaviour of the PageRank algorithm during this computation \cite{becchetti2006distribution}. The resulting PPR values obtained from this process are stored in a dictionary and returned as the function's output. It is  effectively used as a meaningful feature for further analysis.

\begin{algorithm}
  \caption{Compute Personalised PageRank}\label{alg:ppr}
  \DontPrintSemicolon
  \KwIn{$train\_data$: DataFrame containing training data}
  \KwOut{$ppr$: Dictionary of personalised PageRank values}

   Group data by 'creditoraccountnumberhash' and compute 'avg\_label\_creditor'\;
  Group $train\_data$ by 'debtoraccountnumberhash' and compute 'avg\_label\_debtor'\;

      \For{$debtor$ \textbf{in} debtor labels}{
    Retrieve debtor account number and average label\;
    Add entry to $personalisation$ with account number as key and average label as value\;
  }

  \For{$creditor$ \textbf{in} creditor labels}{
    Retrieve creditor account number and average label\;
    Add entry to $personalisation$ with account number as key and average label as value\;
  }

  Compute PPR using the $reference\_graph\_nx$, with a damping factor of 0.85 and the $personalisation$ dictionary\;
\end{algorithm}

\subsubsection{Performance metrics}

The model's performance is assessed using key metrics from a confusion matrix: AUC, Accuracy, Recall, and Precision. AUC, derived from the ROC curve, gauges the model's ability to distinguish between fraudulent and legitimate cases \cite{muschelli2020roc}. Higher AUC indicates superior classification and better discrimination \cite{narkhede2018understanding}.

Accuracy assesses the classifier's correct classification of data points \cite{AI}. Precision and Recall are crucial metrics in FD. Precision gauges accurately predicted positives, while Recall quantifies correctly identified actual positives \cite{c3AI}.

These metrics offer insights into model effectiveness, showcasing an inverse relationship. Recall is vital in FD, reducing missed fraud cases at the cost of accepting some false positives. Precision calculates true positives among predicted positives. In FD, prioritising the reduction of false negatives makes Recall more important than Precision.

\subsubsection{Feature Importance}

Feature importance is evaluated in this work by calculating the absolute coefficient values of the model. These coefficients represent the weights assigned to each feature, indicating their impact on the target variable. By taking the absolute values of these coefficients, the magnitude of the influence of each feature is determined, regardless of its directionality. Comparing the absolute coefficient values allows for ranking features based on their importance. Features with higher absolute coefficients are considered more influential in predicting the outcome and indicate stronger associations with the target variable.

This assessment of feature importance provides valuable insights into the relevance and significance of different features in the model \cite{hooker2018evaluating}. It helps identify the key drivers of the predicted outcomes and enhances the understanding of the underlying relationships between the features and the target variable. This enables the prioritisation and focuses on the most influential variables in the analysis, leading to a more comprehensive interpretation of the model's results \cite{hooker2018evaluating}.

\subsubsection{Population Stability Index}

Ensuring feature stability is crucial for maintaining model integrity, particularly in financial risk modelling. The Population Stability Index (PSI) is a widely adopted metric to evaluate the consistency of feature distributions \cite{min2018behavior}. 

In this paper, the PSI is computed by comparing the distributions of features between the training set and the test set. User-defined bins are created to capture the numerical values present in the actual distribution. The percentage of data from the actual and expected distributions falling into these bins is then compared. When identical, the number of units falling into the bins remains the same. The PSI considers the total number of samples in the actual distribution $N_{a}$ and the expected distribution $N_{e}$, as well as the counts of units from the source distribution $N_{a,bi}$ and the target distribution $N_{e,bi}$ that fall into each bin \cite{khademi2023model}.

To calculate the PSI, the proportion of units from the source distribution falling into each bin is obtained by dividing $N_{a,bi}$ by $N_{a}$, resulting in $q_{a,bi}$. Similarly, the proportion of units from the expected distribution falling into each bin is computed by dividing $N_{e,bi}$ by $N_{e}$, yielding ${q_{e,bi}}$ \cite{khademi2023model}.

The PSI is then calculated as follows:

\begin{equation}
    PSI(A,E) = \sum\limits_{i} (q_{a,bi}-q_{e,bi})\times ln \frac{q_{a,bi}}{q_{e,bi}}
\end{equation}

Low PSI values suggest that the patterns and characteristics captured by these features exhibit consistent and reliable behaviour across different datasets. Such stability contributes to the model's overall robustness and generalisation capabilities, implying that the predictive power encapsulated by these features is likely to persist in real-world scenarios.

\section{Results}
\label{sec:results}

\subsection{Model Performance}

The comparative analysis of the results is presented in Table \ref{table: RF results}, providing valuable insights into the performance evaluation of the models. Notably, the inclusion of PPR demonstrates a significant improvement for the enhanced model ($LR\_ppr$), surpassing the baseline model ($LR\_base$) by 2\% in terms of the AUC metric. This improvement highlights the substantial impact of incorporating the network feature on the model's predictive power. Further, when Accuracy, Precision, and Recall values are rounded to two decimal places, they show high similarity. This suggests that marginal differences between the two models have a negligible impact on performance evaluation.

\begin{figure*}
    \begin{subfigure}{0.45\textwidth}
        \includegraphics[width=\textwidth]{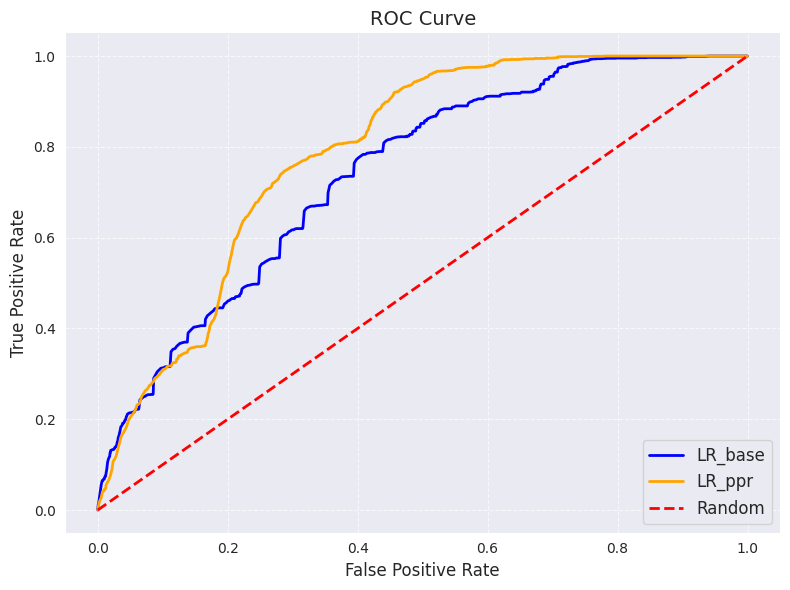}
        \caption{AUC Curve Comparison.}

        \label{fig:ROC comparison}
    \end{subfigure}
    \hspace*{\fill} 
    \begin{subfigure}{0.45\textwidth}
      \includegraphics[width=\textwidth]{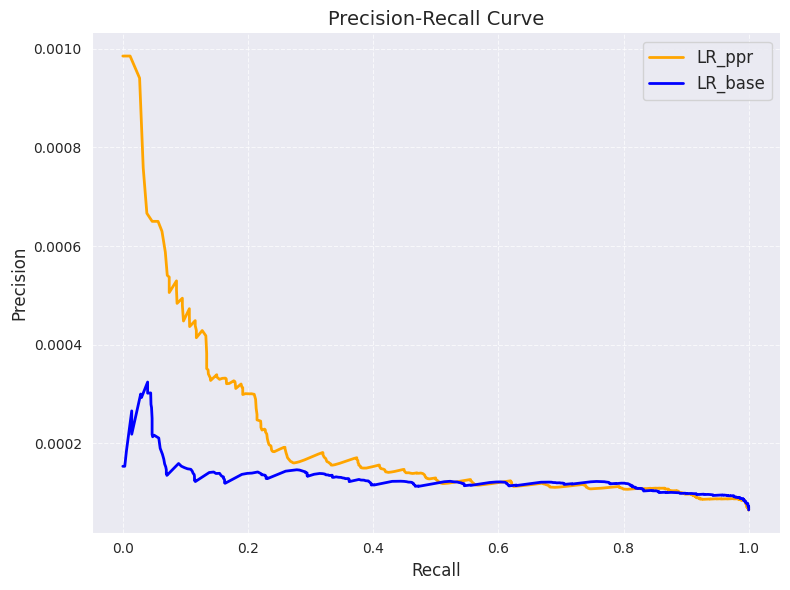}
      \caption{Precision Recall Curve Comparison.}
      \label{fig:ROC comparison2}
    \end{subfigure}
    \caption{ LR\_base (baseline model built on traditional features) vs LR\_ppr (enhanced model built on both traditional and graph feature ppr)} \label{fig:1}
\end{figure*}

Moreover, the performance gain is visually depicted in Figure \ref{fig:1}. In Figure \ref{fig:ROC comparison}, the ROC curves of both models are juxtaposed, showcasing the superior performance of $LR\_ppr$. Its curve exhibits a higher TPR at various FPR thresholds, indicating its improved capability to classify positive instances while minimising FPs accurately.

Furthermore, Figure \ref{fig:ROC comparison2} provides additional insights into the model's performance and compares the Precision-Recall curves of the two models. It becomes apparent that the enhanced model exceeds the base model across various recall levels, resulting in higher precision values for corresponding recall values compared to the base model.

\begin{table}[H]
\footnotesize
\centering
\begin{tabular}{ |p{1cm}|p{1.2cm}|p{0.6cm}|p{1.3cm}|p{0.9cm}|p{1.2cm}|}
 \hline
 \multicolumn{1}{|c}{}&\multicolumn{5}{|c|}{\textbf{Metric}} \\
 \hline
 \textbf{Models}  &  \textbf{Features} &  \textbf{AUC} & \textbf{Accuracy} &  \textbf{Recall} & \textbf{Precision} \\
 \hline
 LR\_base & 6 & 0.74 & 0.99 & 0.99 & 0.99 \\
 LR\_ppr & 7 & 0.76 & 0.99 & 0.99 & 0.99 \\

 \hline
 \end{tabular}
\caption{Experimental Results Logistic Regression.} \label{table: RF results}
\end{table}

\subsection{Interpretability}

\begin{figure}[H]
  \centering
  \includegraphics[width=245pt]{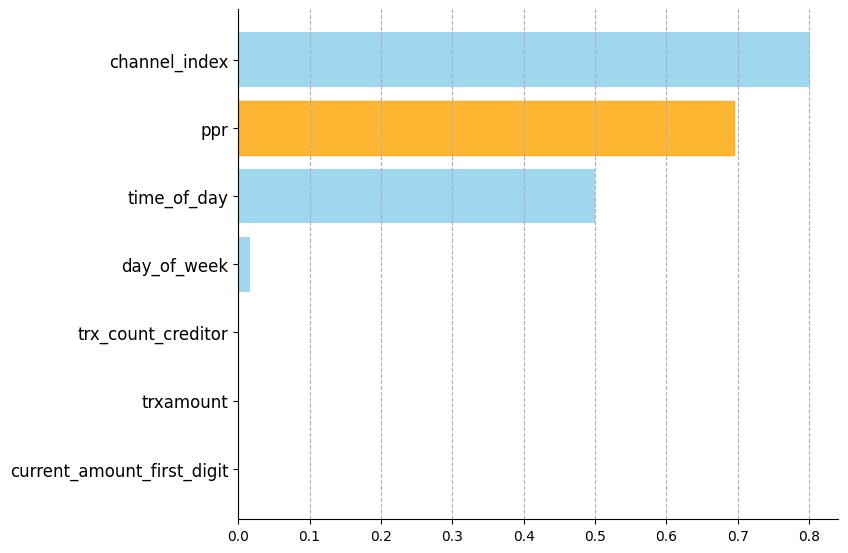}
  \caption{Feature Contribution List: orange for graph features v.s. blue for baseline features.}
  \label{fig:importance}
\end{figure}

Information gain derived from logistic regression analysis is employed to gain valuable insights into the contribution of different features. The resulting feature importance list, depicted in Figure \ref{fig:importance}, uncovers key findings about the significance of various features. Notably, the feature \texttt{channel\_index} emerges as the most influential, exhibiting a remarkable feature importance score of $0.8$. This metric measures the influence of the channel through which the transaction occurred and determines the overall risk greatly. Another noteworthy observation is the pivotal role played by the $ppr$ feature, representing the network characteristics. It demonstrates substantial importance with a score of nearly $0.7$. Additionally, the features \texttt{time\_of\_day} and \texttt{day\_of\_week} also contribute, with importance scores of $0.5$ and $0.01$, respectively. On the other hand, transaction-related features such as \texttt{current\_ amount} and \texttt{current\_amount\_first\_digit} rank considerably lower in terms of feature contribution, displaying zero importance.

\subsubsection{Feature Stability}

In this analysis, feature stability was examined to assess the consistency of feature distributions between the training set and the test set. The results in Table \ref{table:psi} showcase each feature's calculated PSI values.

The analysis findings revealed that all PSI values fall below the predetermined threshold of 0.05. This indicates that the distributions of the features remain stable across the two datasets, with no substantial shifts or discrepancies observed.

\begin{table}[H]   
\footnotesize
\centering
\caption{PSI Values}  
\begin{tabular}{| l |p{1.5cm}|}  
\hline
 \multicolumn{2}{|c|}{\textbf{Overview of PSI Values}} \\
 \hline
\textbf{Feature} & \textbf{PSI} \\ \hline  
\texttt{current\_amount} &  0.01 \\   
\texttt{current\_amount\_first\_digit} & 0.0 \\
\texttt{channel\_index} & $1.62e^{-07}$   \\ 
\texttt{trx\_count\_creditor} & $7.87e^{-05}$ \\
\texttt{day\_of\_week} & $2.35e^{-07}$  \\   
\texttt{time\_of\_day} & $1.17e^{-07}$ \\
\texttt{ppr} & $3.59e^{-08}$ \\
\hline
\end{tabular}  
\label{table:psi}
\end{table}

\section{Discussion}

\label{sec:discussion}

%\subsection{Findings}

The results of the comparative analysis provide valuable insights into the performance evaluation of the models, highlighting three key findings which are listed as follows.

\textbf{Finding (1):} The performance evaluation of the models, as measured by the AUC metric and the Precision-Recall curves, highlights a noteworthy improvement when PPR is included in the $LR\_ppr$ model. Specifically, $LR\_ppr$ surpasses $LR\_base$ by 2\% in terms of AUC, indicating enhanced predictive power. This finding aligns with existing literature \cite{molloy2017graph}, highlighting the positive impact of integrating network-based features, including PageRank, in improving the differentiation between fraudulent and non-fraudulent transactions. The study by \citet{molloy2017graph} supports our $LR\_ppr$ model, which incorporates network features to understand transaction relationships better and enhance fraud identification. Our findings corroborate and extend previous research, emphasising the effectiveness of network-based features for improving the model's ability to discern between fraudulent and non-fraudulent transactions.

\textbf{Finding (2):} The analysis of feature importance reveals intriguing insights into the significance of different features. The feature \texttt{channel\_index} emerges as the most influential, with a remarkable feature importance score of $0.8$. Additionally, the network feature \texttt{ppr} demonstrates substantial importance, scoring nearly $0.7$.

The analysis of feature importance reveals intriguing insights into the significance of different features. This finding aligns with \cite{min2018behavior}, which evaluated feature importance based on information gain. Their research highlighted the prominence of graph features in ranking high as strong signals of associated risk, similar to our findings, however, individual features ranked similarly or even higher.

The findings from \cite{min2018behavior} support our observation that certain features, such as \texttt{channel\_index} and \texttt{ppr}, play a critical role in determining overall risk. These features capture contextual information and network characteristics, enabling accurate risk assessments. Additionally, the study suggests that other individual features, potentially transaction-specific attributes, may possess comparable or superior discriminative power in our specific context. Hence, our finding aligns with existing literature, emphasising the importance of graph features like \texttt{ppr} in risk assessment and acknowledging the significance of individual features e.g., \texttt{channel\_index}. The combination of various feature types contributes to the overall model performance, as well as it can help distinguish risk levels.

\textbf{Finding (3):} Examining feature stability aims to assess the consistency of feature distributions between the training and test datasets. The results indicate that all PSI values, which measure the shift in feature distributions, fall below the predetermined threshold of $0.05$.

This Finding signifies that the distributions of the features remain stable across the two datasets, indicating no substantial shifts or discrepancies. The consistency in feature distributions implies that changes do not influence the models' performance in the underlying data distribution between the training and test sets. Therefore, the insights derived from the analysis can be considered reliable and generalised.

In summary, the discussion of the results encompasses three key findings: the performance improvement indicated by the AUC metric and Precision-Recall curves when incorporating PPR, the positive influence of \texttt{ppr} on risk determination, and the feature stability observed across the training and test datasets. These findings collectively contribute to a deeper understanding of the models' performance and highlight the importance of incorporating network characteristics in risk assessment.

\subsubsection{Limitations}

While this work offers valuable insights into a specific graph feature's impact on the model's performance, it is crucial to recognise limitations that may affect the interpretation and generalisability of the findings.

This paper is limited by using a specific graph network primarily obtained from ING. While valuable insights can be gained, the findings may lack generalisability due to the network's specificity to ING and its context. This restricts the applicability of the paper's conclusions to other financial contexts, where unique patterns, connections, and risk profiles may impact model performance and interpretation 

A second limitation of this research lies in its narrow focus on personalised PageRank as a single exposure network feature. This work may not fully encompass the potential impact of other relevant exposure features by exclusively examining one feature. Including additional self-discovered exposure network features could significantly contribute to a more comprehensive understanding of their collective influence on the model's performance. However, considering the defined scope and objectives of this research, the independent development of multiple new features would surpass the intended boundaries of this paper.

\section{Concluding Remarks}
\label{sec:conclusion}

%\subsection{Conclusion}

Integrating personalised PageRank (\emph{ppr}) notably enhances FD model performance in a logistic regression model with six baseline features. The \emph{ppr} feature shows a 2\% increase in AUC, enhancing predictive power and discrimination between fraudulent and non-fraudulent transactions. It excels in accurately classifying positive instances while minimising false positives. 
%Key features, such as \texttt{channel\_index} and \emph{ppr}, highlight the importance of transaction channel and network characteristics in fraud risk assessment. 
Stability analysis demonstrates consistent distributions between training and test datasets, affirming the model's reliability and generalisability.

%In conclusion, this paper underscores the substantial benefits of integrating \emph{ppr} in FD models, contributing valuable insights to the FD literature and highlighting the importance of leveraging network characteristics for enhanced FD methods.

%\subsection{Future Work}

Future research includes merging data from diverse financial networks to address current limitations and create a more representative dataset for exposure feature analysis. Extensions such as  feature extraction methods, such as node or graph embedding can be interesting to investigate. 

%can effectively represent graph-structured data for supervised learning.Sensitivity analyses with alternative network datasets are advised to assess the observed effects' robustness. Expanding beyond an exclusive focus on PPR for exposure network features is recommended. This involves exploring various centrality measures, graph-based metrics, and novel algorithms tailored to capture exposure patterns. Comparative analyses of diverse exposure features for anomaly detection would yield valuable insights. Additionally, exploring Neural Network models is suggested for enhanced performance. Investigating feature extraction methods, like node embedding, can effectively represent graph-structured data for supervised learning. For feature importance, employing methods like SHAP, LIME, and SAGE is recommended, allowing a deeper understanding of individual features' impact on model predictions and enhancing FD model interpretability.

\bigskip

\bibliography{aaai24}

\end{document}